\documentclass[12pt,preprint]{aastex}

\shorttitle{r-Process and Black Hole Formation}
\shortauthors{Boyd et al.}

\begin{document}


\title{The r-Process in Metal Poor Stars and Black Hole Formation}


\author{R.N. Boyd}
\affil{Physics and Life Sciences, L-050, 
Lawrence Livermore National Laboratory, 
Livermore, CA 94550, USA}
\email{boyd11@llnl.gov}

\author{M.A. Famiano}
\affil{Dept. of Physics and Joint Institute for Nuclear Astrophysics, Western Michigan University, 
1903 W. Michigan Avenue, Kalamazoo, MI 49008-5252, USA}

\author{B.S. Meyer}
\affil{Dept. of Physics and Astronomy, 
Clemson University, 118 Kinard Laboratory, 
Clemson, SC 29634-0978 USA}

\author{Y. Motizuki}
\affil{RIKEN Nishina Center, Hirosawa 2-1, Wako, 351-0198 Japan}

\author{T. Kajino\altaffilmark{1}}
\affil{National Astronomical 
Observatory of Japan, 
2-21-1 Mitaka, Tokyo, 181-8588, Japan}

\and

\author{I.U. Roederer}
\affil{Carnegie Observatories, 813 Santa 
Barbara Street, Pasadena, CA 91101, USA}

\altaffiltext{1}{Dept. of Astronomy, 
Graduate School of Science; Univ. of Tokyo, 7-3-1, 
Hongo, Bunkyo-ku, Tokyo 113-0033, Japan}


\begin{abstract}
Nucleosynthesis of heavy nuclei in metal-poor stars is generally thought to occur via the 
r-process because the r-process is a primary process that would have operated early in the Galaxy's history.
This idea is strongly supported by the fact that
the abundance pattern in many metal-poor stars matches well the inferred Solar r-process
abundance pattern in the mass range between the second and third r-process abundance peaks.
Nevertheless, a significant number of metal-poor stars do not share this standard 
r-process template. In this Letter we suggest that the nuclides observed in many of these stars are 
produced by the r-process, but that it is prevented from running to completion in more 
massive stars by collapse to black holes before the r-process is completed, 
creating a ``truncated r-process,'' or ``tr-process.'' 
We find that the observed fraction of tr-process stars is qualitatively what one would 
expect from the initial mass function, and that an apparent sharp truncation observed at around 
mass 160 could result from a combination of collapses to black holes and the difficulty of 
observing the higher mass rare earths. We test the tr-process hypothesis with r-process 
calculations that are terminated before all r-process trajectories have been ejected.
We find qualitative agreement between observation and theory when black hole collapse and observational 
realities are taken into account.
\end{abstract}
\keywords{stars: Population II --- nucleosynthesis --- black hole physics}
\section{Introduction}
The r-process has been understood for many years to (1) synthesize 
half the nuclides heavier than iron, (2) synthesize all of the nuclides 
heavier than $^{209}$Bi, and (3) be primary in the sense that its nucleosynthesis 
does not appear to depend on preexisting nuclides
\citep{b2fh,woosley94,wallerstein97,farouqi09}. In this latter 
context, its production of the heavier r-process nuclides in many metal poor stars 
appears to be very uniform \citep{sneden03a}, and it also produces relative 
r-process abundances that are essentially the same as those found in more modern stars. 
A standard interpretation of the r-process suggests that it occurs in the neutrino 
wind that emanates from core-collapse supernovae 
\citep{woosley94, takahashi94,farouqi09}, although that interpretation is not 
without its issues \citep{qian96, meyer98, hudepohl10a,hudepohl10b, fischer10, roberts10}. 
Even with these difficulties, 
though, there are a sufficient number of uncertainties, e.g., the possibility of 
sterile neutrinos \citep{mclaughlin99, mclaughlin03}, that this r-process site might turn out to be viable.

At the same time, other metal poor stars exhibit some of the features of r-process 
nucleosynthesis, but their abundances represent a surprisingly poor match with 
the ``standard'' abundance template \citep{aoki00, honda07, roederer10a}, 
perhaps most frequently identified as that observed in 
CS-22892-052 \citep{sneden03a}. A recent paper \citep{roederer10a}
has summarized the situation that exists for metal-poor stars. The data in that paper seem
to suggest that a distribution of r-process abundances exists in metal poor stars, 
with some resembling the standard abundance set, but with a significant fraction 
of stars having abundances that do not match the standard template. These latter 
stars appear to favor the lighter r-process nuclides at varying levels, and many 
seem to have abundance patterns that terminate around Dy, that is, around mass 160.

In this Letter we point out that the abundance patterns observed for the stars that 
do not fit the standard r-process template could be produced by stars that are 
sufficiently massive that their core collapse first produce neutron stars
but that the infall that occurs following the formation of the neutron star subsequently 
results in collapse to a black hole - so called ``fallback supernovae.'' 
This class of stars spans a mass range from roughly 25 to 40 solar 
masses \citep{heger92} for low-metallicity stars. 
When the neutron star collapses to a black 
hole the r-process enrichment of the interstellar medium would cease, 
terminating either when the r-processed regions were swallowed by the black hole, 
or when the electron antineutrinos fell below the event horizon \citep{sasaqui05}.
Thus, this truncated r-process, or tr-process, nucleosynthesis would terminate at 
different stages of that process,
depending on the precise time at which the black hole prevented
further r-process production or emission of nuclides into the interstellar medium. 
We suggest that the delayed collapse to the black hole, 
combined with another effect, namely, the difficulties in observing the higher mass 
rare earths, could produce the cutoff in the r-process distributions observed around 160 u. 

The general scenario we envision merely assumes that mass elements that predominantly produce the
lighter r-process nuclei are ejected before the mass elements that produce the heavier r-process nuclides.
Any setting within a core-collapse supernova that satisfies this condition would allow for
a truncated r-process.  To evaluate this idea quantitatively, however, we apply the
neutrino-driven wind model for the r-process.  Although, as noted above, this model
is not without issues; it is plausible and well discussed in the literature and makes a good setting for discussing
the truncated r-process.

\section{Neutrino Wind Model of the r-Process}
In the neutrino wind scenario of the r-process, neutrinos from the nascent neutron
star heat material at the neutron star surface and drive it away in a wind.  As a wind
element leaves the neutron star surface, it expands and cools.  Nucleons in the
wind element assemble into heavier nuclei, which serve as the seeds for the subsequent
r-process \citep{woosley94}. Importantly, the abundance pattern for the seed nuclei 
is well-described by a quasi-equilibrium, which, for the entropy and neutron richness
characteristic of neutrino-driven wind envirnoments, has a relatively sharp
abundance peak near A$\sim$100.  As the temperature in the wind element falls further,
charged-particle reactions freeze out.  If a high neutron density accompanies
the seed nuclei, neutron captures and beta decays successively promote
the seed nuclei to higher mass.  This upwards flow in mass slows
at the neutron closed shells at 82 and then 126 neutrons, and 
thereby produces the r-process abundance peaks at A$\sim$130 and 195 u 
\citep{woosley94, wanajo02}.

For what follows, it is essential to note that the neutrino-driven wind leaves the
star over a period of several seconds.  In the standard scenario
\citep{meyer92, woosley94, takahashi94}, the wind elements that leave the star early
have lower entropy and a lower degree of neutron richness.  As the neutron star
evolves, the wind elements that leave later have higher entropy and a greater
degree of neutron richness.  It is in these latter wind elements that the heaviest
nuclei are made.  The lower-entropy, less neutron rich wind elements that leave
the neutron star earlier produce ligher nuclei.  If the neutron star collapses to a
black hole after the lower-entropy elements leave the star but before the higher-entropy
do, then the r-process has been truncated, and the abundance pattern will be dominated
by lighter r-process nuclei.

Current supernova models do not naturally produce the parameters 
required to produce a successful r-process, most notably the entropy, but
also the requisite neutron richness.  While it may in fact be the case that
core-collapse supernovae are not the site of the r-process, it may also be that
current supernova models are not sufficiently detailed to provide an adequate
description of the r-process conditions in core-collapse events.
For example, results from multidimensional hydrodynamics calculations suggest that the 
instabilities resulting from those calculations may ultimately be shown to produce 
the entropy required for making the heaviest r-process nuclei \citep{burrows06}.
Interestingly, these instabilities may generate the neutron star
kicks \citep{burrows06, guilet10} that give rise to high space velocities of some
pulsars \citep{arzou02}.
It will also be necessary
to include all of the detailed neutrino physics \citep{burrows02, roberts10, liebendoerfer05}
to adequately characterize the neutron richness in matter expelled from the supernova.

Future work will determine whether more advanced core-collapse supernova models will provide the
conditions needed for the r-process.  For the purposes of the present paper,
we will simply assume that
neutrino-driven winds in core-collapse supernovae are at least one of the sites of
r-process nucleosynthesis.

\section{Model Calculations}
To study the truncated r-process in the neutrino-driven wind,
we applied the basic idea behind the study of \citet{woosley94}, 
which assumed that the r-process occurred in the neutrino wind from a core collapse supernova. In that 
study, a succession of 40 ``trajectories'' (that is, thin shell wind elements), all assumed to 
have originated deep within the (assumed spherically symmetric) star, but 
having initially different conditions of density, 
temperature, entropy, and electron fraction, were emitted into the 
interstellar medium, thus contributing 
to the total r-process nucleosynthesis. The bubble 
was evolving in time, so that the conditions under which the individual trajectories were 
processed changed with the identity 
of the trajectory. We assumed that the different 
trajectories were emitted 
from the star successively, but ceased to be emitted when the collapse to the black hole 
occurred. This would be consistent with \citet{woosley94}, who assumed successive 
emissions of the trajectories to generate what turned out to be a good representation of 
the Solar  r-process abundances.

\begin{figure}
\plotone{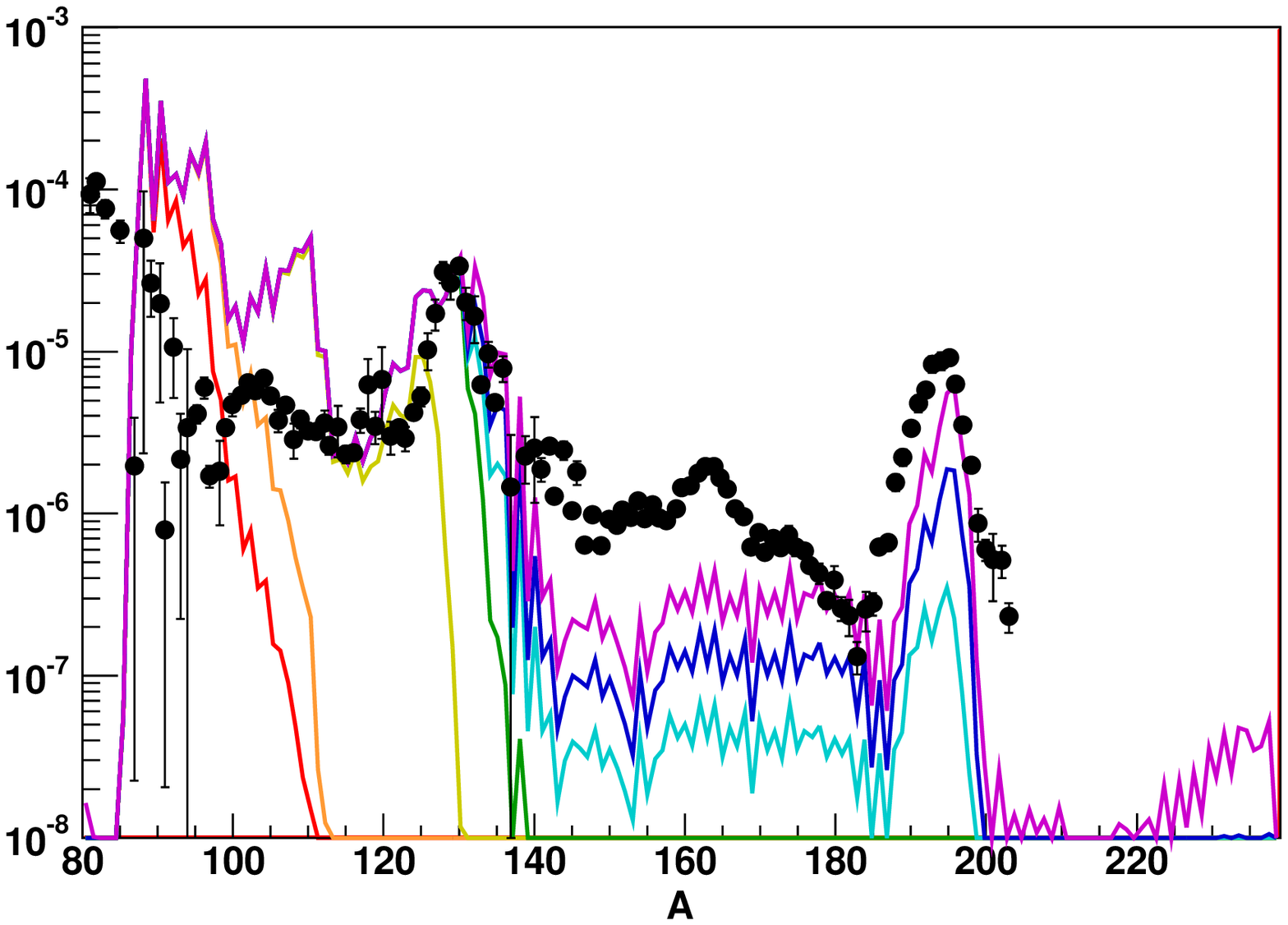}
\caption{r-Process calculations using the trajectories from \citet{woosley94}, but 
summing the results from trajectory number 24 to some later trajectory, 
as was done in \citet{woosley94}. The successive curves include 
trajectory 24, 24 through 26, 24 through 28, 
24 through 30, 24 through 31, 24 through 32, and 
24 through 40, where the emission times of the latest trajectory in each sum 
are as indicated. The sums to higher number trajectories make both the mass 130 and 195 
r-process peaks, and the sum through trajectory 40 comes closest to representing the 
Solar r-process abundances \citep{kappeler89}, shown as the dots.\label{woo_traj}}
\end{figure}

For our r-process calculations, we used a network code  
based on libnucnet, a library of C codes for storing and managing nuclear reaction 
networks \citep{meyer07}.  The nuclear and reaction data for the calculations were
taken from the JINA reaclib database \citep{cyburt10}.
We performed calculations for trajectories 
24 - 40 in the \citet{woosley94} hydrodynamics model.
For each trajectory, reaction network calculations were 
performed for T$_9<$2.5 using initial abundances derived from the \citet{woosley94}
results.  Our calculations were simplified by assuming an initial abundance of massive 
nuclei from a single nucleus heavy seed with a mass A equal to the average mass at 
T$_9$=2.5 and an atomic number derived from the average mass number, the Y$_e$, and 
the neutron and alpha mass fractions at T$_9$=2.5 in \citet{woosley94}.  We justify
this by noting that the heavy-nuclide distribution is typically sharply peaked near T$_9$=2.5.
An adiabatic expansion was assumed for each trajectory in the nucleosynthesis, with entropy
constant within a trajectory but varying between trajectories, again consistent with 
the approach of \citet{woosley94} for times at which the temperature T$_9<$2.5. 
The material was taken to expand at constant velocity on a timescale consistent with
that derived from \citet{woosley94}.
For each trajectory the calculation was continued 
until neutron capture reactions had ceased and the abundance distribution versus
mass number had frozen out. Our representation of the 
full r-process abundances, shown as the dots in Figure \ref{woo_traj}, 
is not as good as that 
achieved by \citet{woosley94},
but our calculations do produce the basic features of the 
full r-process, namely the mass 130 and 195u peaks.
%

The simulation of \citet{woosley94} that produced a good r-process representation 
began with trajectory number 24 and summed the nucleosynthesis yields from the remaining 
16 elements since all of these trajectories would have been ejected into the interstellar
medium. We also began with trajectory 24, and 
performed a mass weighted sum of the nucleosynthesis from the trajectories up to some 
higher number trajectory to observe the resulting nucleosynthesis when the trajectories 
beyond our maximum trajectory were assumed to be consumed by the collapse to a black 
hole. 
This produced a result similar to that of \citet{woosley94} when all trajectories 
from 24 to 40 were included. The results are shown in Figure \ref{woo_traj}. There it can be seen that 
truncating the r-process at increasing trajectory number does terminate the r-process nucleosynthesis 
at increasingly higher nuclear mass. Note that although the curve representing trajectories 
24 through 31 does reach the mass 195u peak, the abundance produced in that calculation 
is nearly two orders of magnitude below that of the full r-process, which would render the
higher mass nuclides difficult to observe. 
The abundances for that calculation would therefore appear observationally 
to terminate at a mass of about 140 u.

\begin{figure}
\includegraphics[scale=0.9]{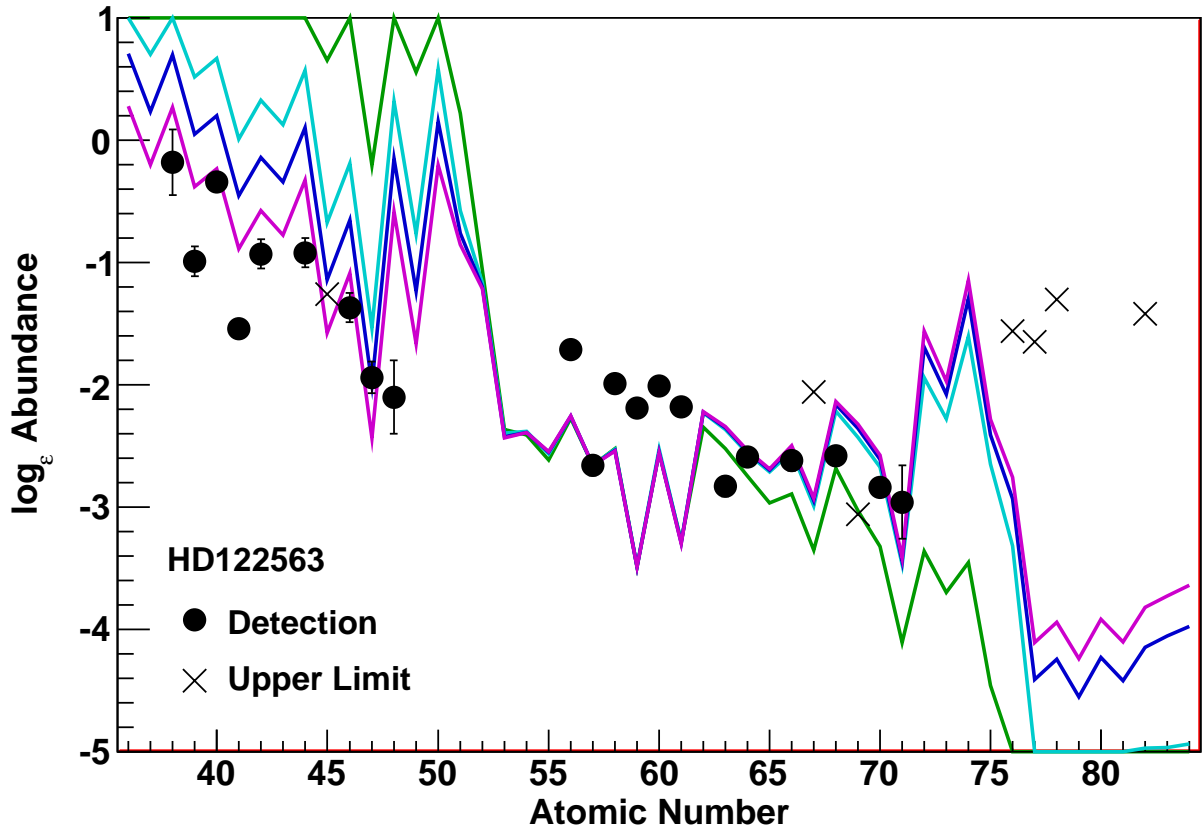}
\caption{Comparison of tr-process predictions, using the formalism adapted from 
\citet{woosley94}, for four mass cuts.  These are cuts below shell 30 (green), 31 
(light blue), 32 (dark blue), and 40 (purple) with abundances in the metal-poor star 
HD 122563. Observational data are from \citet{honda06} and Roederer et al. (2009, 2010b).  
The predictions are scaled to the La abundance (Z=57).
\label{ele_abun}}
\end{figure}

In Figure \ref{ele_abun}  we compare several tr-process calculations with the derived elemental 
abundance pattern in the metal poor halo star HD 122563. This low-metallicity star ([Fe/H] 
= -2.7) is deficient in the heavy neutron-capture elements (Ba and heavier) relative to the 
light neutron-capture elements (Sr through Cd) when compared with the scaled Solar 
r-process pattern. The HD 122563 abundances are a very poor match to the scaled Solar 
s-process pattern. Its abundance pattern matches the scaled Solar r-process pattern 
better up to an atomic mass of about 70 (mass of $\sim$174 u), but even this fit is 
unsatisfying \citep{sneden83, honda06}. Stars like 
HD 122563 may be candidates for enrichment by the tr-process. Figure \ref{ele_abun} demonstrates 
that the tr-process predictions, while far from a perfect match to the individual 
abundances, can reproduce the relative downward trend in abundance with increasing 
atomic number seen in some metal poor stars. More exploration of tr-process calculations 
is obviously needed, but the general trend is encouraging.

\section{Probability of Occurence of tr-Process Stars}
In principle, a test of the tr-process model would be provided by a large set of 
data for metal poor stars that spans the masses of the nuclides produced in the 
r-process from mass $\sim$80 to the heaviest observable nuclides that the r-process 
synthesizes,  usually thorium. Unfortunately this is challenging due to the difficulty 
of observing the higher mass rare-earth nuclides, starting at mass $\sim$160. Thus all 
r-process events that terminate at masses between 160 and lead would appear to terminate 
at mass 160, producing the effect observed by \citet{roederer10a}. 

Any standard initial mass function (IMF) reflects the fact that less-massive stars are 
more numerous than more-massive stars.  Assuming that stars from 8 to 40 solar 
masses may produce comparable amounts of r-process material, we can use the IMF 
to estimate the fraction of stars whose r-process may be truncated by collapse to a 
black hole.  We adopt the \citet{salpeter54} IMF, for which dN/dm $\propto$ m$^{-2.35}$ 
for massive stars, where m is the stellar mass in units of the solar mass, and N is the 
number of stars of a given mass per unit volume.  The ratio of the number of stars that 
would be expected to collapse to black holes (25-40 solar masses) to those expected to 
collapse to neutron stars (8-25 solar masses) is 0.13 in a well-sampled IMF.

While a number of r-process rich stars (here taken to mean stars with [Eu/Fe] $>$ +1.0) 
have been discovered and studied in detail in the last two decades, these stars constitute 
a relatively minor fraction of all metal-poor stars.  More unbiased samples 
\citep{mcwilliam95, barklem05} find that stars with [Eu/Fe] $>$ +1.0 comprise $<$10\% 
of all stars with [Fe/H] $<$ -2.0.  Figure 11 of \citet{roederer10a} suggests that stars 
with [Eu/Fe] $<$ 0 are candidates for enrichment by a truncated r-process.  Difficulties 
in detecting Eu in metal-poor stars with [Eu/Fe] $<$ 0 make it even more challenging to 
estimate how numerous these stars are.  Ba is more easily detected and may be used to 
represent Eu and other heavy elements in stars lacking s-process enrichment.  From the 
Ba abundances in the large survey of \citet{barklem05} we estimate a lower limit of 
$\sim$55\% of metal-poor stars as candidates for enrichment by a truncated r-process. 
This is significantly higher than the 13\% derived above 
assuming a Salpeter IMF, yet Figure \ref{ele_abun} demonstrates that a 
tr-process may be one way to produce the heavy nuclides observed in 
some metal-poor stars. The discrepancy in these fractions could indicate that additional 
nucleosynthesis channels, such as those proposed by, e.g., 
\citet{qian08}, \citet{farouqi09}, \citet{wanajo11}, or \citet{nakamura11}, 
together with the tr-process may contribute to the chemical enrichment of the early 
Galaxy. 
Given that massive stars with short lifetimes will dominate the chemical 
enrichment at early times, though, the tr-process may very well have been a major source of 
heavy nuclei at these epochs. 

\section{Other r-Process Issues}
As a possible additional benefit of the tr-process, we note that the stars that were 
truncated at mass points 28 or less produced no nuclides in the mass 130 peak or beyond. 
This would have the effect of boosting the yields of the lighter r-process nuclides 
relative to the heavier ones, filling in the mass 110 to 120 u region. It has been known 
for some time that r-process calculations that produce a mass 195 u peak underproduce 
the nuclides in the 110-120 mass region. The r-process nuclides that were even 
lighter than those were underproduced by an even larger factor, but this would 
also be consistent with distributed cutoffs of the r-process nucleosynthesis. Thus, 
again guided by the IMF, but along with the mass dependent cutoff times, the tr-process 
might potentially solve this troublesome aspect of current r-process analyses 
\citep{kratz07,farouqi09}.
Conclusions in this regard, however, must 
await more detailed analyses.

An obvious test of the tr-process model could occur from renewed effort to observe the 
higher mass rare earth elements in very metal poor stars with low [Eu/Fe] ratios. If 
elements in that mass region can be observed, and the tr-process is correct, additional 
observations would be expected to map out the black hole collapse times over that region. 
While no doubt challenging, a relationship between observations and black hole collapse 
times would be of great interest.
\acknowledgments
RNB's work has been supported under the auspices of the Lawrence Livermore 
National Security, LLC (LLNL) under Contract No. DE-AC52-07NA27344; MF's 
by National Science Foundation grant PHY-0855013; BSM's by NASA grant 
NNX10AH78G; YM's by the NEXT Program of JSPS and CSTP (GR098); 
TK's by Grants-in-Aid for Scientific Research of JSPS (20244035), 
Scientific Research on Innovative Area of MEXT (20105004), and Heiwa Nakajima 
Foundation; and IUR's by the Carnegie Institution of Washington through the 
Carnegie Observatories Fellowship. This is document LLNL-JRNL-491647.

\clearpage


\clearpage


\end{document}